\documentclass[superscriptaddress,aps,prl,preprintnumbers,nofootinbib,preprint,11pt]{revtex4}
\usepackage[utf8]{inputenc}
\usepackage[citecolor=blue,urlcolor=blue,unicode=true,pdfusetitle,
bookmarks=true,bookmarksnumbered=true,bookmarksopen=true,bookmarksopenlevel=3,
breaklinks=false,pdfborder={0 0 1},backref=false,colorlinks=true]{hyperref}
\usepackage{bm}
\usepackage{latexsym}
\usepackage{dcolumn,xcolor}
\usepackage[normalem]{ulem}
\usepackage{amsmath,amsfonts,amssymb}
\usepackage{graphicx,epsfig}
\usepackage{soul}
\usepackage{ulem}
\usepackage{amsthm}
\usepackage{color,float} 

\usepackage{amssymb}
\usepackage{braket}
\usepackage{physics}

\usepackage{tikz}

\begin{document}
\title{ Consequences of Undecidability in Physics on the Theory of Everything}

\author{Mir Faizal\footnote{Email: \href{mailto:mirfaizalmir@gmail.com}{mirfaizalmir@gmail.com}}}
\affiliation{ \scriptsize{Irving K. Barber School of Arts and Sciences, 
  University of British Columbia - Okanagan, Kelowna,
British Columbia V1V 1V7, Canada.}}
\affiliation{\scriptsize{Canadian Quantum Research Center, 204-3002 32 Ave, Vernon, BC V1T 2L7, Canada.}}
\affiliation{ \scriptsize{Department of Mathematical Sciences, Durham University, Upper Mountjoy, Stockton Road, Durham DH1 3LE, UK.}}
\affiliation{\scriptsize{Faculty of Sciences, Hasselt University, Agoralaan Gebouw D, Diepenbeek, 3590, Belgium.}}
\author{Lawrence M. Krauss\footnote{Email: \href{mailto:lawrence@originsproject.org}{lawrence@originsproject.org}}}
\affiliation{\scriptsize{Origin Project Foundation, Phoenix, AZ 85018, USA.}}
\author{Arshid Shabir\footnote{Email: \href{mailto:aslone186@gmail.com}{aslone186@gmail.com}}}
\affiliation{\scriptsize{Canadian Quantum Research Center, 204-3002 32 Ave, Vernon, BC V1T 2L7, Canada.}}
\author{Francesco Marino\footnote{Email: \href{mailto:francesco.marino@ino.cnr.it}{francesco.marino@ino.cnr.it}}}
\affiliation{\scriptsize{CNR-Istituto Nazionale di Ottica and INFN, Via Sansone 1, I-50019 Sesto Fiorentino (FI), Italy.}}

\date{\today}

\begin{abstract}
General relativity treats spacetime as dynamical and exhibits its breakdown at singularities. This failure is interpreted as evidence that quantum gravity is not a theory formulated  {within} spacetime; instead, it must explain the very  {emergence} of spacetime from deeper quantum degrees of freedom, thereby resolving singularities. Quantum gravity is therefore envisaged as an axiomatic structure, and algorithmic calculations acting on these axioms are expected to generate spacetime. However, Gödel’s incompleteness theorems, Tarski’s undefinability theorem, and Chaitin’s information-theoretic incompleteness establish intrinsic limits on any such algorithmic programme. Together, these results imply that a wholly algorithmic “Theory of Everything’’ is impossible: certain facets of reality will remain computationally undecidable and can be accessed only through non-algorithmic understanding. We formalize this by constructing a “Meta-Theory of Everything’’ grounded in non-algorithmic understanding, showing how it can account for undecidable phenomena and demonstrating that the breakdown of computational descriptions of nature does not entail a breakdown of science. Because any putative simulation of the universe would itself be algorithmic, this framework also implies that the universe cannot be a simulation.

 \end{abstract}

\maketitle

 Physics has journeyed from classical  tangible “stuff’’ to ever deeper layers of abstraction.  In Newtonian mechanics reality consists of point-like masses tracing deterministic trajectories in an immutable Euclidean space with a universal time parameter   \cite{Landau1976}.  This picture sufficed for celestial mechanics and terrestrial dynamics, yet its very foundations, including the separability of space and time and the notion of absolute simultaneity, were overturned by Einstein’s special relativity.  By welding space and time into a single Lorentzian continuum, special relativity replaced Newton’s rigid arena with an  observer-dependent spacetime geometry whose interval, not time or space separately, is invariant   \cite{Rindler1977}.

Quantum mechanics introduced a second conceptual revolution: even with a fixed spacetime backdrop, the microscopic world resists classical deterministic descriptions.  Wave functions evolve unitarily, but measurement outcomes are inherently probabilistic, encoded in the Born rule and constrained by complementarity and uncertainty principles   \cite{Sakurai2017}.  When the relativistic requirement of locality is imposed on quantum theory, particles cease to be fundamental.  Instead, quantum field theory (QFT) elevates fields to primary status; “particles’’ emerge from those fields via creation and annihilation operators acting on the vacuum state   \cite{Srednicki2007}.
Here the vacuum is itself a seething medium.  Time-dependent boundary conditions in superconducting wave-guides emulate moving mirrors and catalyse the dynamical Casimir effect, producing real particles from vacuum fluctuations   \cite{Wilson2011}.  Likewise, an accelerated observer perceives the Minkowski vacuum as a thermal bath via the so-called the Unruh effect, emphasizing that particle content is observer-dependent rather than absolute   \cite{Crispino2008}.  These phenomena confirm  QFT: what we call a particle is contingent on both the quantum state of quantum fields and even the kinematics of the detector. Thus, particles moving in spacetime become   a contingent structure, yet spacetime remains fundamental and fixed.

All these theories presuppose a fixed background spacetime. General relativity (GR), by contrast, is a theory of spacetime itself. It accurately describes phenomena from Mercury’s perihelion precession to the direct detection of gravitational waves   \cite{Einstein1915, Abbott2016}. Nevertheless, GR predicts curvature singularities at the center of black holes and at the big bang, where the spacetime  description of reality breaks down   \cite{Penrose1965, Hawking1970}. Singular behavior of this sort is not unique to gravity; it signals the breakdown of any effective model once its underlying degrees of freedom are pushed beyond their domain of validity   \cite{Arnold92, Berry2023}. Classical fluid discontinuities, for example, correspond to curvature singularities of an acoustic metric and are smoothed out in a full quantum-hydrodynamic treatment   \cite{Faccio2016, Braunstein:2023jpo}.

Thus, it is expected that curvature singularities in GR will also be removed in a full quantum theory of gravity. These singularities do not indicate a breakdown of physics, but the breakdown of a spacetime description of nature. Instead, it is presumed the physics of a quantum theory of gravity will  not break down, even in such extreme conditions. 
Candidate quantum gravity frameworks likewise remove curvature singularities. Loop quantum cosmology replaces the big bang singularity with a big bounce   \cite{Bojowald2001, Ashtekar2006}, while the fuzzball paradigm in string theory substitutes extended microstate geometries for point-like singularity at the center of  black holes    \cite{Mathur2005, Mathur2008}. More broadly, both loop quantum gravity and string theory depict spacetime as emergent: spin-foam models build it from discrete quantum structures   \cite{Perez2013}, and the doubled-geometry formalism of double field theory introduces T-folds whose transition functions involve T-duality rather than ordinary diffeomorphisms, showing that classical spacetime may fail to be  well defined at some points   \cite{Hohm2010, Hull2005}.

These insights resonate with Wheeler’s “it from bit’’ programme and its modern  versions  in both string theory   \cite{Jafferis:2022crx, VanRaamsdonk:2020ydg} and loop quantum gravity   \cite{Makela:2019vgf}, which propose that information is more  fundamental than  physical reality consisting of spacetime and quantum fields defined on it   \cite{Wheeler1990}. Singularities in classical models then mark precisely those regions where the informational degrees of freedom can no longer be captured by a spacetime geometry.
Although the emergent “it’’ spacetime with its quantum fields fails at singularities, one might hope that the underlying “bit’’, a complete quantum-gravity theory, could be formulated as a consistent, computable “theory of everything.’’ However, we now argue that  that such a purely algorithmic formulation is unattainable.

As we do not have a fully consistent theory of quantum gravity, several different  axiomatic systems  have been proposed to model  quantum gravity   \cite{Witten:1985cc,Ziaeepour:2021ubo,Faizal2024,bombelli1987spacetime,Majid:2017bul,DAriano:2016njq,Arsiwalla:2021eao}.  
In all these programs, it is assumed 
a candidate theory of quantum gravity is encoded as a {computational} formal system
\begin{equation} {\;
    \mathcal{F}_{QG}
    \;=\;
    \{\mathcal{L}_{QG},\Sigma_{QG},\mathcal{R}_{\mathrm{alg}} \} 
  \;}.
  \end{equation}
Here,  $\mathcal{L}_{QG}$ a first-order language whose non-logical symbols denote quantum states, fields, curvature, causal relations, {etc.}
 $\Sigma_{QG}=\{A_1,A_2,\dots\}$ is a {finite} (or at least recursively-enumerable) set of closed $\mathcal{L}_{QG}$-sentences embodying the fundamental physical principles.
 $\mathcal{R}_{\mathrm{alg}}$ the standard, {effective} rules of inference  used for computations.
They operationalise “algorithmic calculations’’; we write
$         \Sigma_{QG}\;\vdash_{\mathrm{alg}}\;\varphi       \quad\Longleftrightarrow\quad
        \varphi\text{ is derivable from }\Sigma_{QG}\text{ via }\mathcal{R}_{\mathrm{alg}}.
      $
Crucially, spacetime is  {not} a primitive backdrop but a theorem-level construct emergent inside models of $\mathcal{F}_{QG}$.
Concrete mechanisms for which such geometry can emerge include dynamics in  string theory   \cite{Seiberg2006,Polchinski1998}, entanglement in holography   \cite{Jafferis:2022crx,VanRaamsdonk:2020ydg}, and spin-network dynamics in LQG   \cite{Perez2013,Rovelli2004,Makela:2019vgf}.


Any viable $\mathcal{F}_{QG}$ must meet four intertwined criteria:
  {Effective axiomatizability;} The number of axioms in $\Sigma_{QG}$  are finite. This  ensures that proofs are well-posed. In fact, it is expected  that spacetime can be algorithmically generated from this, and so it has to be computationally well defined   \cite{Faizal2023,Braunstein:2023jpo}.
  {Arithmetic expressiveness;} $\mathcal{L}_{QG}$  can internally model the natural numbers with their basic operations. This is important as quantum gravity should reproduce calculations  used  for amplitudes, curvature scalars, entropy, etc in appropriate limits.   Both string theory   \cite{Polchinski1998,Green1987} and LQG    \cite{Rovelli2004,Thiemann2007} satisfy this by reproducing GR and QM in appropriate limits.  {Internal consistency;} no $\Sigma_{QG}\vdash_{\mathrm{alg}}\bot$.  Strings secure this via anomaly cancellation   \cite{Green1984,Polchinski1998}; LQG via an anomaly-free constraint algebra   \cite{Ashtekar1986,Rovelli2004}.
  {Empirical completeness;} predictive all physical phenomena  from the Planck scale to cosmology, and even resolves  singularities.

The axiom set $\Sigma_{QG}$ is finite, arithmetically expressive  and  consistent.  As a result, Gödel’s incompleteness theorems apply   \cite{Godel1931,Smith2007}.
 Here, we consider the algorithmic core of quantum gravity as a finite, consistent and arithmetically expressive formal system $\mathcal{F}_{QG}=\bigl(\mathcal{L}_{QG},\Sigma_{QG},\mathcal{R}_{\mathrm{alg}}\bigr)$.  Its deductive closure is the recursively enumerable set of theorems $\operatorname{Th}(\mathcal{F}_{QG})=\{\varphi\in\mathcal{L}_{QG}\mid\Sigma_{QG}\vdash_{\mathcal{R}_{\mathrm{alg}}}\varphi\}$, while the semantically true sentences are $\operatorname{True}(\mathcal{F}_{QG})=\{\varphi\in\mathcal{L}_{QG}\mid\mathbb N\models\varphi\}$. Thus,  Gödel’s first incompleteness theorem asserts the strict containment $\operatorname{Th}(\mathcal{F}_{QG})\subsetneq\operatorname{True}(\mathcal{F}_{QG})$  \cite{Godel1931,Smith2007}, guaranteeing the existence of well‑formed $\mathcal{L}_{QG}$-statements that are  {true but unprovable} within the algorithmic machinery of $\mathcal{F}_{QG}$.  Physically these Gödel sentences correspond to empirically meaningful facts—e.g., specific black‑hole microstates—that elude any finite, rule‑based derivation.  Gödel’s second theorem deepens the impasse: the self‑referential consistency statement $\operatorname{Con}(\mathcal{F}_{QG})\equiv\neg\operatorname{Prov}_{\Sigma_{QG}}(\bot)$ cannot itself be proved by $\mathcal{F}_{QG}$ without contradiction  \cite{Godel1931,Smith2007}.  A purely computational theory of everything would therefore not be able to establish its own internal soundness.
Tarski’s undefinability theorem further bars the construction of an internal truth predicate $\mathsf{Truth}(x)\in\mathcal{L}_{QG}$ obeying $\Sigma_{QG}\vdash_{\mathcal{R}_{\mathrm{alg}}}[\mathsf{Truth}(\ulcorner\varphi\urcorner)\leftrightarrow\varphi]$ for all $\varphi$  \cite{Tarski1933,Tarski1983,Faizal20241}. So, a truth predicate for quantum gravity cannot be defined within the theory itself.   Finally, Chaitin’s information‑theoretic incompleteness establishes a constant $K_{\mathcal{F}_{QG}}$ such that any sentence $S$ with prefix‑free Kolmogorov complexity $K(S)>K_{\mathcal{F}_{QG}}$ is undecidable in $\mathcal{F}_{QG}$  \cite{chaitin1975theory,Chaitin2004,kritchman2010surprise}.  This bound caps the epistemic reach of algorithmic deduction by declaring ultra‑complex statements—inevitable in high‑energy quantum gravity—formally inaccessible.

Together, the Gödel–Tarski–Chaitin triad delineates an insurmountable frontier for any strictly computable framework.  To attain a genuinely complete and self‑justifying theory of quantum gravity one must augment $\mathcal{F}_{QG}$ with non‑algorithmic resources—an external truth predicate  axioms, or other meta‑logical mechanisms—that transcend recursive enumeration while remaining empirically consonant with physics at the Planck scale.
Although these limits restrict what can be known computationally,  the Lucas–Penrose argument shows that non-algorithmic understanding can access truths beyond formal proofs    \cite{lucas1961minds,Penrose2011,Penrose1990-PENTNM,hameroff2014consciousness,lucas_penrose_2023}. Purely algorithmic deduction is therefore insufficient for a complete foundational account \cite{Faizal:2025gip}.

To transcend these computations limitations, we adjoin an external truth predicate $T(x)$ and a non-effective inference mechanism $\mathcal{R}_{\mathrm{nonalg}}$, enlarging the formal apparatus to
\begin{equation}
  \mathcal{M}_{\mathrm{ToE}}
  =
\{
      \mathcal{L}_{QG}\!\cup\!\{T\},
      \Sigma_{QG}\!\cup\!\Sigma_T,
      \mathcal{R}_{\mathrm{alg}}\!\cup\!\mathcal{R}_{\mathrm{nonalg}}
   \}.
\end{equation}
Here $\Sigma_T$ is an external, non-recursively-enumerable set of axioms about $T$.  We write $\Sigma_T\vdash_{\mathrm{nonalg}}\varphi$ precisely when $T(\ulcorner\varphi\urcorner)\in\Sigma_T$.
The   external truth predicate  axioms obey four intertwined conditions.  (S1)  {Soundness for $\mathcal{F}_{QG}$}: whenever $T(\ulcorner\varphi\urcorner)$ is an axiom, $\varphi$ holds in every model of the base theory.  (S2)  {Reflective completeness}: if $\varphi$ is algorithmically derivable from $\Sigma_{QG}$, then the implication $\varphi\!\rightarrow\!T(\ulcorner\varphi\urcorner)$ itself belongs to $\Sigma_T$.  (S3)  {Modus-ponens closure}: $T$ respects logical consequence, for $T(\ulcorner\varphi\!\rightarrow\!\psi\urcorner)$ together with $T(\ulcorner\varphi\urcorner)$ entails $T(\ulcorner\psi\urcorner)$.  (S4)  {Trans-algorithmicity}: the induced theory $\operatorname{Th}_T=\{\varphi\mid T(\ulcorner\varphi\urcorner)\in\Sigma_T\}$ is not recursively enumerable; sentences of arbitrarily high Kolmogorov complexity can still be $T$-true, exceeding the information bound $K_{\mathcal{F}_{QG}}$.

With these properties the   external truth predicate  certifies every Gödel sentence of $\mathcal{F}_{QG}$ and can single out, for instance, concrete black-hole microstates that elude all algorithmic searches, thereby side-stepping the information-loss puzzle and illuminating Planck-scale dynamics.  The non-algorithmic understanding encoded by $\mathcal{R}_{\mathrm{nonalg}}$ and $\Sigma_T$ thus supplies conceptual resources inaccessible to purely computational physics.

For clarity of notation: $\Sigma_{QG}$ is the computable  axiom set; $\mathcal{R}_{\mathrm{alg}}$ comprises the standard, effective inference rules; $\mathcal{R}_{\mathrm{nonalg}}$ is the non-effective   external truth predicate  rule that certifies $T$-truths; $\mathcal{F}_{QG}=\{\mathcal{L}_{QG},\Sigma_{QG},\mathcal{R}_{\mathrm{alg}}\}$ denotes the computational core; and $\mathcal{M}_{\mathrm{ToE}}=\{\mathcal{L}_{QG}\cup\{T\},\Sigma_{QG}\cup\Sigma_T,\mathcal{R}_{\mathrm{alg}}\cup\mathcal{R}_{\mathrm{nonalg}}\}$ denotes the full meta-theory that weds algorithmic deduction to an   external truth predicate.

Crucially, the appearance of undecidable phenomena in physics already offers  empirical backing for $\mathcal{M}_{\mathrm{ToE}}$.  Whenever an experiment or exact model realises a property whose truth value provably eludes every recursive procedure, that property functions as a concrete witness to the  truth predicate $T(x)$ operating within the fabric of the universe itself.  Far from being a purely philosophical embellishment, $\mathcal{M}_{\mathrm{ToE}}$ thus emerges as a structural necessity forced upon us by the physics of undecidable observables.  Working at the deepest layer of description, $\mathcal{M}_{\mathrm{ToE}}$ fuses algorithmic and non-algorithmic modes of reasoning into a single coherent architecture, providing the semantic closure that a purely formal system $\mathcal{F}_{QG}$ cannot reach on its own.  In this enriched setting, quantum measurements,  Planck-scale processes, quantum-gravitational amplitudes and cosmological initial conditions might all become accessible to principled yet non-computable inference, ensuring that no physically meaningful truth is left outside the scope of theoretical understanding. Just as Riemannian geometry, which describes general relativity, or gauge theories, which describe various interactions of the Standard Model, are each actualized in nature, this truth predicate 
 \(T(x) \) would also be actualized in nature.

The logical limitations reviewed above bear directly on several open questions in quantum gravity, beginning with the black-hole information paradox   \cite{Almheiri2021}.  If the microstates responsible for the Bekenstein–Hawking entropy live at Planckian scales, where smooth geometry breaks down, Chaitin’s incompleteness theorem suggests that their detailed structure may forever lie beyond algorithmic derivation.  In such circumstances, classical spacetime must re-emerge through a collective, effectively thermal, behaviour of microscopic degrees of freedom.  Yet deciding whether a given many-body system thermalises is itself algorithmically undecidable   \cite{Shiraishi2021}.  Here $\mathcal{M}_{\mathrm{ToE}}$ becomes indispensable: by adjoining the  external truth predicate $T(x)$ that certifies physically admissible yet uncomputable properties, the meta-theory legitimizes the passage from undecidable Planck-scale microphysics to the macroscopic notion of spacetime thermalization.

Thermalization already plays a central role in leading quantum-gravity models.  In AdS/CFT, bulk perturbations relax into black-hole horizons whose thermodynamic parameters are sharply defined   \cite{Chesler:2009cy}; in the fuzzball paradigm, an ensemble of horizonless microstate geometries reproduces the Hawking spectrum   \cite{Mathur:2005zp}; and in LQG, coarse-graining drives discrete quantum geometries toward a classical continuum phase   \cite{dittrich2020coarse}.  Because thermalisation is undecidable in the general many-body setting   \cite{Shiraishi2021}, each route from Planck-scale physics to smooth spacetime must contain steps that transcend algorithmic control.  The non-algorithmic scaffold provided by $\mathcal{M}_{\mathrm{ToE}}$ supplies precisely the logical footing required to keep such trans-computational steps consistent.

Computational undecidability likewise shadows other structural questions in many-body physics and hence in quantum gravity.  No algorithm can decide in full generality whether a local quantum Hamiltonian is gapped or gapless   \cite{cubitt2015undecidability}; the proof embeds Turing’s halting problem   \cite{turing1936computable}, which links back to Chaitin’s theorem   \cite{li1997introduction}.  Entire renormalization-group flows can behave uncomputably   \cite{watson2022uncomputably}, even though RG ideas underpin string-theoretic beta-functions   \cite{Callan:1985ia}, background-independent flows in LQG   \cite{Steinhaus:2018} and continuum-limit programmes such as asymptotic safety and causal dynamical triangulations   \cite{Litim:2004,Ambjorn:2005db}.  If generic RG trajectories defy algorithmic prediction, then translating fundamental quantum-gravity data into classical spacetime observables again lies beyond finite computation.  By embedding these flows into $\mathcal{M}_{\mathrm{ToE}}$, one places them under a broader logical umbrella where non-computational criteria rooted in $T(x)$ can still certify physical viability.

Related undecidable sectors abound.  Key properties of tensor networks ubiquitous in holography   \cite{hayden2016holographic} and LQG   \cite{Dittrich:2011zh} are formally uncomputable   \cite{kliesch2014matrix}.  Deducing supersymmetry breaking in certain two-dimensional theories is undecidable   \cite{tachikawa2023undecidable}, influencing model building in string theory   \cite{Green1984}.  Phase diagrams of engineered spin models encode uncomputable problems   \cite{bausch2021uncomputability}, and the mathematical kinship between such systems and LQG kinematics   \cite{Feller:2015} hints at analogous intractabilities in the full phase structure of loop gravity.  Each undecidable domain slots naturally into $\mathcal{M}_{\mathrm{ToE}}$, which extends explanatory reach beyond algorithmic barriers while maintaining logical coherence through its  external truth predicate  axioms.

{These technical results respect rather than undermine the  principle of sufficient reason    \cite{amijee2021principle,leibniz1996discourse}.  The core demand of that principle is that every true fact must be grounded in an adequate explanation. This forms the basis of science.   Gödel incompleteness, Tarski undefinability, and Chaitin bounds do not negate this demand; they merely show that  “adequate explanation’’ is broader than “derivable by a finite, mechanical procedure.’’   In other words, the existence of true but unprovable $\mathcal{L}_{QG}$-sentences does not imply that those facts lack reasons, but only that their reasons need not be encoded  syntactically  within any recursively enumerable axiom set.  The semantic  external truth predicate  $T$ introduced above models such non-algorithmic grounding: it certifies truth directly at the level of the underlying mathematical structure, thereby supplying sufficient reasons that transcend the deductive reach of $\Sigma_{QG}$.  Thus, far from conflicting with the principle of sufficient reason, the logical limits on computation  affirm  it by revealing that explanatory resources extend beyond formal proof theory. So, a breakdown of computational explanations does not imply a breakdown of science. 

Many undecidable statements encountered in physics ultimately trace back to the halting problem   \cite{bennett1990undecidable}, yet non-algorithmic understanding can still apprehend such truths   \cite{Stewart1991}.  The Lucas–Penrose proposal that human cognition surpasses formal computation   \cite{lucas1961minds,Penrose2011,Penrose1990-PENTNM,hameroff2014consciousness,lucas_penrose_2023} finds a mathematical expression in $\mathcal{M}_{\mathrm{ToE}}$, whose external truth predicate $T(x)$ certifies propositions that no algorithmic verifier can capture.  In line with the orchestrated objective-reduction (OR) proposal, they claim that human observers can have  a truth predicate because cognitive processes exploit quantum collapse,  which is produced by the truth predicate of quantum gravity   \cite{hameroff2014consciousness}. This is why they argue that human mathematicians can apprehend Gödelian truths, whereas computers cannot.  

Non-algorithmic reasoning already supplements GR through the Novikov self-consistency principle   \cite{Friedman1990,novikov1989}, which imposes a global logical constraint on spacetimes with closed timelike curves.  By housing such meta-principles in $\mathcal{M}_{\mathrm{ToE}}$ one side-steps Gödelian obstructions that would cripple a purely formal $\mathcal{F}_{QG}$. 
As quantum logic is itself undecidable   \cite{vandenNest2008measurement,lloyd1993quantum}, any proper wave-function–collapse mechanism must operate outside the algorithmic domain of the quantum mechanics. So, such dynamics naturally reside in the non-algorithmic  $\mathcal{M}_{\mathrm{ToE}}$. Gravitationally induced objective-collapse proposals can therefore be interpreted as concrete instantiations of the $\mathcal{M}_{\mathrm{ToE}}$ action on quantum states   \cite{Penrose1996,Diosi1987}. Here,  the meta-layer supplies a non-algorithmic  gravity-triggered collapse that is not derivable from  $\Sigma_{QG}$, but is nonetheless well-defined at the semantic level. A key advantage of using  objective-collapse models might be cosmological: it could offer a explanation of the quantum-to-classical transition in cosmology, thereby addressing the measurement problem in quantum  cosmology   \cite{Gaona-Reyes:2024qcc}.

 A growing survey confirms that undecidability permeates diverse areas of physics   \cite{peraleseceiza2024undecidabilityphysicsreview}.  These examples jointly reinforce the proposition that a quantum-gravity rooted solely in computation can be neither complete nor consistent, whereas augmenting it with the non-algorithmic resources encoded in $\mathcal{M}_{\mathrm{ToE}}$ could restore explanatory power without losing logical soundness.

The claim that our universe is itself a computer simulation has been advanced in several forms, from Bostrom’s statistical “trilemma’’   \cite{Bostrom2003} to more recent analyses by Chalmers   \cite{Chalmers2019} and Deutsch   \cite{Deutsch2016}.  These proposals assume that every physical truth is reducible to the output of a finite algorithm executed on a sufficiently powerful substrate.  Yet this assumption tacitly identifies the full physical theory with its computable slice \( \mathcal{F}_{QG} \).
 
Our framework separates the computable fragment \( \mathcal{F}_{QG} \) from the non-algorithmic meta-layer \( \mathcal{M}_{\mathrm{ToE}} \).  Because \( \mathcal{M}_{\mathrm{ToE}} \) contains an external truth predicate \( T(x) \) that by construction escapes formal verification, any finite algorithm can at best emulate \( \mathcal{F}_{QG} \) while  {systematically omitting} the meta-theoretic truths enforced by \( T(x) \).  Consequently, no simulation could in principle reproduce what would otherwise be the full underyling structure of the physics of our universe.  Our analysis instead suggests that genuine physical reality embeds non-computational content that cannot be instantiated on a Turing-equivalent device. Since it is impossible to simulate a complete and consistent universe, our universe is definitely not a simulation. 
 As the universe is produced by \( \mathcal{M}_{\mathrm{ToE}} \), the simulation hypothesis is logically impossible rather than merely implausible.

The arguments presented here suggest that neither `its' nor `bits' may be sufficient to describe reality. Rather, a deeper description, expressed not in terms of information but in terms of non-algorithmic understanding, is required for a complete and consistent theory of everything.

\section*{Acknowledgments} We would like to thank İzzet Sakallı, Salman Sajad Wani, and Aatif Kaisar Khan
 for useful discussions. We would also like to thank Aatif Kaisar Khan for sharing with us an important paper on undecidability.     Stephen Hawking's discussion on G\"odel’s theorems  and the end of physics motivated the current work.  We would also like to thank Roger Penrose for his exploration of G\"odel’s theorems and the Lucas-Penrose argument, which forms the basis of meta-theoretical perspective based on non-algorithmic understanding. 
\bibliographystyle{utphys}
\bibliography{references}
\end{document}